\documentclass[pra,twocolumn,amsmath,amssymb,superscriptaddress]{revtex4-2}
\usepackage{graphicx,amsmath,relsize,epstopdf,color,mathtools,bm,newtxtext,newtxmath,braket,rotating}
\usepackage[hyphenbreaks]{breakurl}
\usepackage[colorlinks=true,linkcolor=blue,citecolor=blue,urlcolor =blue]{hyperref}
\usepackage[normalem]{ulem}
\usepackage[table,xcdraw]{xcolor}
\usepackage{braket}

\usepackage{soul,xcolor}

\usepackage{siunitx}

\begin{document}

\title{Observation of robust polarization squeezing via the Kerr nonlinearity in an optical fibre}

\author{N.~Kalinin}
\affiliation{Max Planck Institute for the Science of Light, 91058 Erlangen, Germany}
\affiliation{Nonlinear Dynamics and Optics Division, Institute of Applied Physics of the Russian Academy of Sciences, 603950 Nizhny Novgorod, Russia}

\author{T.~Dirmeier}
\affiliation{Max Planck Institute for the Science of Light, 91058 Erlangen, Germany}
\affiliation{Department of Physics, Friedrich-Alexander-Universit\"at Erlangen-N\"urnberg, 91058 Erlangen, Germany}

\author{A.~Sorokin}
\affiliation{Nonlinear Dynamics and Optics Division, Institute of Applied Physics of the Russian Academy of Sciences, 603950 Nizhny Novgorod, Russia}

\author{E.~A.~Anashkina}
\affiliation{Nonlinear Dynamics and Optics Division, Institute of Applied Physics of the Russian Academy of Sciences, 603950 Nizhny Novgorod, Russia}
\affiliation{Advanced School of General and Applied Physics, Lobachevsky State University of Nizhny Novgorod, 603950 Nizhny Novgorod, Russia}

\author{L.~L.~S\'{a}nchez-Soto}
\affiliation{Max Planck Institute for the Science of Light, 91058 Erlangen, Germany}
\affiliation{Departamento de \'Optica, Facultad de F\'{\i}sica,
Universidad Complutense, 28040 Madrid, Spain}

\author{J.~F.~Corney}
\affiliation{School of Mathematics and Physics, University of Queensland, Brisbane, Queensland 4072, Australia}

\author{G.~Leuchs}
\affiliation{Max Planck Institute for the Science of Light, 91058 Erlangen, Germany}
\affiliation{Nonlinear Dynamics and Optics Division, Institute of Applied Physics of the Russian Academy of Sciences, 603950 Nizhny Novgorod, Russia}
\affiliation{Department of Physics, Friedrich-Alexander-Universit\"at Erlangen-N\"urnberg, 91058 Erlangen, Germany}

\author{A.~V.~Andrianov}
\affiliation{Nonlinear Dynamics and Optics Division, Institute of Applied Physics of the Russian Academy of Sciences, 603950 Nizhny Novgorod, Russia}

\begin{abstract}
Squeezed light is one of the resources of photonic quantum technology. Among the various nonlinear interactions capable of generating squeezing, the optical Kerr effect is particularly easy-to-use. A popular venue is to generate polarization squeezing, which is a special self-referencing variant of two-mode squeezing. To date, polarization squeezing generation setups have been very sensitive to fluctuations of external factors and have required careful tuning. In this work, we report on a development of a new all-fibre setup for polarization squeezing generation. The setup consists of passive elements only and is simple, robust, and stable. We obtained more than 5 dB of directly measured squeezing over long periods of time without any need for adjustments. Thus, the new scheme provides a robust and easy to set up way of obtaining squeezed light applicable to different applications. We investigate the impact of pulse duration and pulse power on the degree of squeezing.
\end{abstract}

\maketitle

\section{Introduction}

Squeezed light refers to a unique set of quantum states of the electromagnetic field \cite{Walls:1983tv,Leuchs1986}, playing an important role in photonic quantum technologies \cite{Loudon1987,Lvovsky2014}. 
The term refers to a phase space description and is used whenever the volume of the quantum uncertainty in phase space is squeezed, so that a variable such as a particular field quadrature has a variance smaller than the one of a coherent state.
After the first observation of squeezing \cite{Slusher1985}, many groups studied the various schemes for the generation, the detection and applications of squeezed states of light as reviewed 30 years later \cite{Andersen:2016vd}. Since then the field progressed further e. g. with the demonstration of interferometric sensitivity enhancement in kilometer-sized gravitational wave detectors \cite{Lough2021} and with the proposal \cite{Hamilton_2017} and subsequent experimental demonstrations \cite{Pan_boson_sampl_2020,Madsen2022} of Gaussian Boson sampling --- two important hallmark milestones in different pillars of quantum technology. Quantum sensing with squeezed light continues to strive \cite{Lawrie2019}. The theoretical concept of squeezed states is, however, much older \cite{Dodonov_2002}.
The following two types of squeezed states of a single mode of light are the most common ones: (1) squeezed vacuum states and (2) squeezed coherent states. Note that squeezed vacuum states can be 'bright' containing many photons \cite{Perez:14}. A state squeezed along a line is a Gaussian state, like a coherent state, and is described by a positive valued Wigner function. In some applications, there still has to be a non-Gaussian element. When combining squeezed states with linear optical networks, non-Gaussianity can be achieved by using nonlinear, i. e. non-Gaussian detectors, such as single photon or photon number resolving detectors \cite{Jonsson2019}. This applies to both types of squeezed states, but depending on how many photons such a detector can handle, type (1) squeezed states could be preferable. Note, that there are also nonlinear squeezed states, e. g. squeezed along a bent line, which are non-Gaussian by definition \cite{Brauer:21} and could become a valuable resource.
In principle, any nonlinear optical light matter interaction \cite{BOYD2008xiii} will change the photon statistics, potentially leading to squeezing, but sizeable squeezing can only be observed if at the same time losses are low enough \cite{Andersen:2016vd}. Therefore, a preferred nonlinear interaction is a non-resonant $\chi ^{(2)}$ interaction in a low-loss crystal, which straightforwardly can generate type (1) \cite{Wu1986} and also type (2) \cite{SIZMANN1990138} squeezed light.
However, this method requires a special material class without inversion symmetry as the optically nonlinear material and a phase matching condition has to be fulfilled. Nevertheless, this is probably the generation scheme used most often so far in applications \cite{SCHNABEL20171}. 
A conceptually more simple scheme generating primarily type (2) squeezed light is based on the optical Kerr effect, i.~e. the non-resonant $\chi ^{(3)}$ interaction. Note, that Kerr squeezing can be generated also by a cascaded $\chi ^{(2)}$ interaction \cite{Noirie1997}, which facilitates the observation of seeing the effect for continuous wave light \cite{Singh2019}. Here we report on a novel scheme for generating Kerr polarization squeezing in a robust way.

\section{State of Kerr effect squeezing of light}

All materials show the Kerr effect, including standard telecommunication fibres, which have the additional benefit of having very small optical losses. The catch is that the Kerr effect changes the photon statistics such as to generate type (2) squeezed light introducing amplitude-phase correlations, resulting in a squeezing ellipse in phase space oriented in a skewed direction, neither in the direction of the amplitude nor in that of the phase quadrature. As a result, the detection of Kerr squeezing has been a challenge. Kerr squeezing can be viewed as four-wave-mixing: two photons are annihilated at the pump frequency, i.~e. the carrier frequency, and one photon each are generated symmetrically in an upper and lower side band, not too different from the $\chi ^{(2)}$ scheme, only that there the carrier in between is missing \cite{Schori2002,Bachor2nd,Stiller2014,SCHNABEL20171}. 
In the regime of anomalous dispersion in a silicon dioxide glass wave guide, phases are matched in a wide spectral range  leading to a correspondingly large noise bandwidth of more than a THz \cite{Spalter:98b}.  
As a result, the side band frequency can not only be small e. g. in the radio frequency range but also large, closer to optical frequencies. In any case, when performing direct detection of the resulting field, no effect of this Kerr interaction is to be seen because the number of photons is unchanged. Alternatively, in the single-mode phase space picture, the different amplitudes within the uncertainty of the initial coherent state experience amplitude dependent phase shifts, resulting in the skewed ellipse so that the overall amplitude uncertainty is unchanged. 

The group first studying this generation scheme in an optical fibre with a continuous wave light beam \cite{Shelby1986} used dispersive back reflection from an optical cavity \cite{GALATOLA1991,Villar2008} to adjust the squeezed ellipse orientation such that squeezing was detectable in direct detection \cite{Shelby1986}\footnote{Note, that the first observation of squeezing by Slusher et al. \cite{Slusher1985} also used four-wave mixing, the nonlinear medium being a sodium atomic beam, and optical cavities separated the pump light from the sidebands, so that type (1) squeezing was generated and measured with heterodyne detection properly adjusting the local oscillator phase. Squeezing light with four-wave mixing in atomic vapours has improved greatly since then \cite{McCormick:07,Marino:2009vn}}. 
But they struggled with thermal phase  noise in the fibre due to scattering on the thermally excited acoustic modes of the fibre, for which they coined the expression 'guided acoustic wave Brillouin scattering' (GAWBS). The observed squeezing was minute. Next, the idea came up to use optical soliton pulses instead \cite{Rosenbluh1991,Drummond1993}, which do not disperse and keep their peak power. So, the effective Kerr nonlinearity was considerably higher, allowing one to use shorter fibres, which reduces GAWBS. In addition, this new approach used a symmetric fibre Sagnac interferometer, where one output contains approximately a squeezed vacuum and the other one a bright state which can be used as a local oscillator to probe the squeezing \cite{Rosenbluh1991,Bergman:91,Bergman:94,Fujiwara:09}. Both outputs experience approximately similar GAWBS noise, so that it largely cancels. As it turned out, this scheme
works well if the beam splitting ratio is exactly $50\% /50\%$ over the whole spectrum and the performance quickly deteriorates even when the deviation is only small \cite{Drummond1993}. But yet another route to observable squeezing was found:  photons in different frequency bands typically show correlations. It is only when you sum over them that there is no observable effect of the Kerr interaction. Friberg \cite{Friberg1996} recognized that if one perturbs this balance, it is possible to observe a sub shot noise signal in direct detection, or quantum correlated signals when detecting different frequency bands separately. Of course, again optical soliton pulses were used. Such spectral manipulation is possible with a grating spectrometer, if the frequency spacing is sufficiently large \cite{spalter1998}, or with a highly dispersive interferometer when the frequency spacing is in the radio frequency regime \cite{Huntington2005,Huntington_2002}. This scheme is closely related to parametric amplification with four-wave-mixing \cite{Liu:16,Serkland:99,Sharping:01} and to quantum frequency combs \cite{Menicucci2008,Pinel2012,Gerke2015,Strekalov_2016,Yang2021,Vuckovic2022}.
The fourth route to observable Kerr squeezing in a fibre was to follow the proposal by Kitagawa and Yamamoto \cite{Kitagawa1986} and use an asymmetric Sagnac loop to displace the squeezing ellipse in phase space such that the short axis lines up with the amplitude quadrature \cite{schmitt1998,Krylov:98,Levandovsky:99}. The scheme was recently used for precision transmission measurements \cite{Atkinson2021}. All the Sagnac interferometer set-ups have the common problem that at most locations along the fibre, the counter propagating fields probe the GAWBS  at different times, so some degradation because of GAWBS is persisting. 
Therefore, new schemes were invented with two co-propagating pulses. A first type was still following Kitagawa and Yamamoto but unfolding the Sagnac interferometer leading to single mode squeezed light \cite{Fiorentino2001,Fiorentino:2002}. The second type \cite{Heersink:05,dong2008,corney_2008,Hosaka:15} uses two equally intense pulses emerging from two separate and independent Kerr interactions in a linear polarization preserving fibre, one soliton pulse propagating along the fast and the other one along the slow axis such that the two coincide at the fibre output. But this time one is not trying to interfere the two pulses to produce single mode squeezed light. Rather, the measured noise reduction is now a property of both modes and one describes it as polarization squeezing which is a special variety of the more general concept of two-mode squeezing \cite{Caves1985,Heidmann1987,Reid1988,Caves1991,Chirkin_1993,Lassen2009}. The two light fields are analyzed using the SU(2) description \cite{Yurke1986} and visualizing this system of two modes on the Poincar\'e sphere \cite{Mueller2012}. There is an interesting dynamics of  the Wigner function on the sphere as a result of the Kerr interaction \cite{Rigas_2013,Valtierra2020}. Note, that measuring the Stokes parameters spanning the Poincar\'e sphere again involves interference and it can be done with direct detection without any additional local oscillator. In this sense, one might say that one beam serves as the local oscillator of the other one and vice versa and that this system of two beams is self-referencing. GAWBS is reduced much further because of the co-propagation of the beams. This has been the fibre set-up generating the highest squeezing so far \cite{dong2008}.

\section{All fiber Kerr squeezing}

In this work we present a major modification of this co-propagating scheme, providing greater stability due to an all-fibre passive design. In this novel setup, a polarization squeezed beam is produced when two equally bright pulses --- simultaneously propagating in orthogonal polarization modes of a polarization-maintaining fibre --- are recombined at the exit of the fibre. Each of these initially coherent pulses experiences Kerr squeezing while traveling through the fibre, changing its shape in phase space to a squeezed ellipse. 
When two such squeezed beams are combined in orthogonal modes, a polarization squeezed beam emerges, in which the variance of one of the Stokes parameters is lower than that in a coherent state with the same intensity. We define the quantum Stokes parameters as follows:
\begin{align}
\hat{S}_0 &= \hat{a}_H^{\dagger} \hat{a}_H + \hat{a}_V^{\dagger} \hat{a}_V, \qquad 
\hat{S}_1 = \hat{a}_H^{\dagger} \hat{a}_H - \hat{a}_V^{\dagger} \hat{a}_V, \nonumber \\
& \\
\hat{S}_2 &= \hat{a}_H^{\dagger} \hat{a}_V + \hat{a}_V^{\dagger} \hat{a}_H, \qquad
\hat{S}_3 = i (\hat{a}_V^{\dagger} \hat{a}_H - \hat{a}_H^{\dagger} \hat{a}_V),\nonumber
\end{align}
where $\hat{a}_{H/V}$ and $\hat{a}_{H/V}^{\dagger}$ are annihilation and creation operators in the horizontal ($H$) and vertical ($V$) polarization modes. Thus, $\hat{S}_0$ is proportional to the pulse energy, and the polarization is circular when $\langle \hat{S}_1 \rangle = \langle \hat{S}_2 \rangle = 0$, $\langle \hat{S}_3 \rangle = \langle \hat{S}_0 \rangle$. In the case of circular polarization, the Heisenberg inequality gives the lower limit on the fluctuations of $\hat{S}_1$ and $\hat{S}_2$:
\begin{equation}
    \Delta^2 \hat{S}_1 \Delta^2 \hat{S}_2 \ge | \langle \hat{S}_3 \rangle |^2,
\end{equation}
where $\Delta ^2 \hat{S}_i$ denotes the variance $ \langle \hat{S}_i^2 \rangle  -  \langle \hat{S}_i \rangle ^2$.
The Stokes parameters can be directly measured using polarization splitting optics and two photodetectors. 

\begin{figure}[t]
  \centering
  \includegraphics[width=0.7\columnwidth]{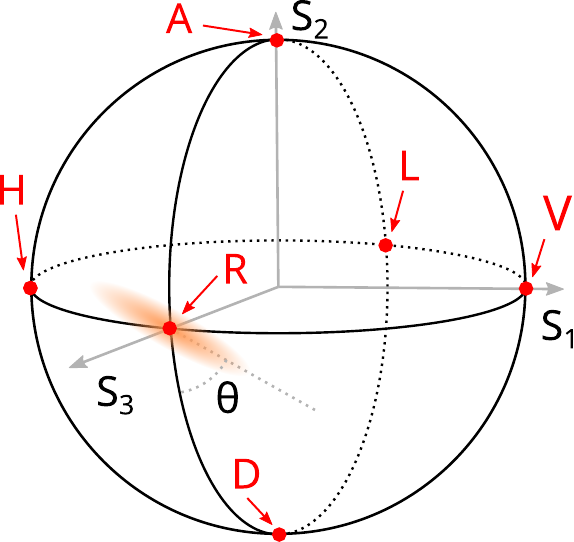}
  \caption{Poincar\'e sphere showing different polarization states (red points and letters), along with the Stokes axes (black letters); the orange ellipse represents the squeezed state of circularly polarized light projected in the sphere with an alignment angle $\theta$.}
  \label{fig:sphere}
\end{figure}

Consider the three dimensional space with axes corresponding to three Stokes parameters $S_1$, $S_2$, $S_3$. Consider a sphere in this space centered at the origin, such sphere is called the Poincar\'e sphere, where each point corresponds to a certain polarization of the beam with fixed power. When two Kerr squeezed beams are combined in orthogonal polarization, we can represent the state of light as an ellipsoid in the Stokes parameters space, which is an ellipse in the projection on the Poincar\'e sphere (Fig. \ref{fig:sphere}). It is easy to position the center of this ellipse at the point corresponding to the circular polarization using standard polarization optics, and after that continuously change its axes alignment angle $\theta$ with a single half-wave plate.
After that, we can measure the $\hat{S}_1$ Stokes parameters fluctuation with a pair of balanced detectors. The fluctuations $\Delta \hat{S}_1$ depend on the alignment angle $\theta$:
\begin{equation}
    \Delta^2 \hat{S}_1^2 = \Delta_s^2 \; \cos^2{\theta} + \Delta_a^2 \; \sin^2{\theta},
    \label{eq:noise_theta}
\end{equation}
where $\Delta_s^2$ denotes the lowest (squeezing) possible variance of $\hat{S}_1$, and the $\Delta_a^2$ denotes the highest (anti squeezing) variance. We see that squeezing is detected when the short axis of the ellipse is parallel to the $S_1$ axis ($\theta = 0$), and $\Delta_s^2 < | \langle \hat{S}_3 \rangle | < \Delta_a^2$. Thus, it’s possible to analyze polarization squeezing without the use of an additional local oscillator, making polarization squeezing promising for applications.

Light propagates in two polarization modes of a polarization-maintaining fibre with different group velocities, thus a birefringence compensation is required to overlap the two pulses in time. In the original setup \cite{corney_2008}, a free-space interferometer was used to match the group delay, and a piezo-mounted mirror in one of the interferometer arms was used to control the relative phases of the two polarizations in a closed loop. In our new setup, we equalize the group delays by splitting the fibre into two equal lengths and rotating one of them by $90^\circ$ around its axis. Thus, one of the two pulses travels first along the slow axis of the fibre, and then along the fast axis, while the other one travels first along the fast axis, and then the slow, resulting in equal group delay of both pulses. The two parts of the fibre are spliced with low losses, allowing the squeezing to aggregate over both parts of the fibre. Moreover, the relative phase drifts of the two pulses in our setup are very little and slow, because both pulses propagate through the same optical path and experience any thermal or mechanical changes equally. This advantage allows us to implement the setup without any active control loop; the polarization of the emerging beam is set to circular with the use of wave plates. Because of that, the setup is very stable.
\begin{figure*}[t]
  \centering
  \includegraphics[width=\linewidth]{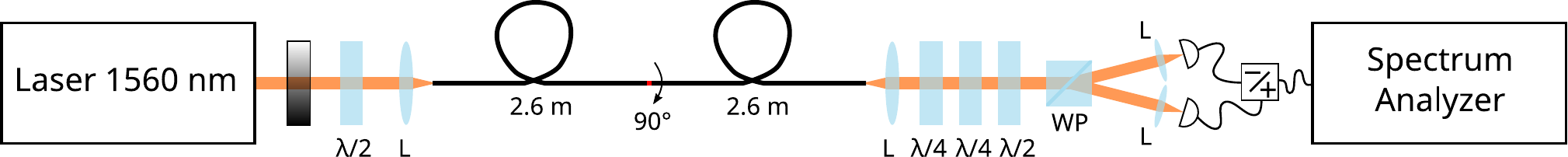}
  \caption{The scheme of the experimental setup. L --- lenses, $\lambda/2$ --- half-wave plates, $\lambda/4$ --- quarter-wave plates, WP --- a Wollaston prism.}
  \label{fig:scheme}
\end{figure*}

\section{Experimental setup}

The scheme of the experimental setup is shown in Fig. \ref{fig:scheme}. Two different setups with the same scheme were implemented (I and II). Femtosecond lasers at 1560 nm were used as the source, with pulse duration of 200 fs (full width half maximum, FWHM, setup I), and adjustable duration of 235-370 fs (FWHM, setup II). The pulses were cosh-shaped temporally, allowing for easy soliton excitation in the fibre. The lasers were checked to be shot noise limited at the frequencies of interest. The repetition rate was 80 MHz with an available pulse energy of up to $1.4$ nJ. A gray filter (setup I) or a half-wave plate and a polarizer (setup II) were installed at the output of the laser to adjust the power, followed by a polarizer and a half-wave plate to adjust the linear polarization of the light launched to the fibre. Thus, both power and polarization of the light can be controlled, which is important to generate polarization squeezing. The polarization was adjusted so that it is $45^\circ$ relative to the axes of the fibre, resulting in two pulses of equal power excited in two polarization modes in the fibre.

The polarization maintaining fibre used is 3M FS-PM-7811 of total length 5.2 m. The fibre was cut into two halves and spliced with a $90^\circ$ rotation of the second half around its axis so that the slow axis of the first half is aligned with the fast axis of the second half and vice versa. The splicing losses were estimated to be $4\%$. After the fibre, the two pulses had an arbitrary relative phase, meaning that the polarization could be anywhere between the diagonal ($45^\circ$ to the fibre axis) and circular, so we used two quarter-wave plates to make the polarization of the beam circular. After that, a half-wave plate was used to rotate the uncertainty ellipse without changing the polarization of the light so that the minor axis of the ellipse is aligned with the H-V axis of the Poincar\'e sphere. 
The half-wave plate flips the clockwise circular polarisation to counter-clockwise and vice versa, but it also changes the orientation of the squeezing ellipse depending on the orientation of the wave plate with respect to the fibre axes. Finally, the beam was split with a Wollaston prism into vertically and horizontally polarized components and detected with a pair of balanced detectors.

The signals from the detectors were low-pass-filtered to attenuate the high-amplitude signal at the 80 MHz pulse repetition rate. After that, the sum or the difference of the signals were fed to an electronic spectrum analyzer (ESA, Agilent E4411B (setup I), E4401B (setup II)). The noise power was detected at a radio frequency sideband of 15 MHz with the resolution bandwidth set to 100 kHz and a video bandwidth of 30 Hz. Thus, when the sum of the signals was detected at the ESA, effectively the variance of the power or the $\hat{S}_0$ Stokes component was measured. When the difference was detected, the variance of the $\hat{S}_1$ Stokes component was measured. All reported measurements were corrected for the electronic noise of $-103.5$ dBm measured in the absence of the light beam.

\section{Results}

\begin{figure*}[t]
  \centering
  \includegraphics[width=0.8\linewidth]{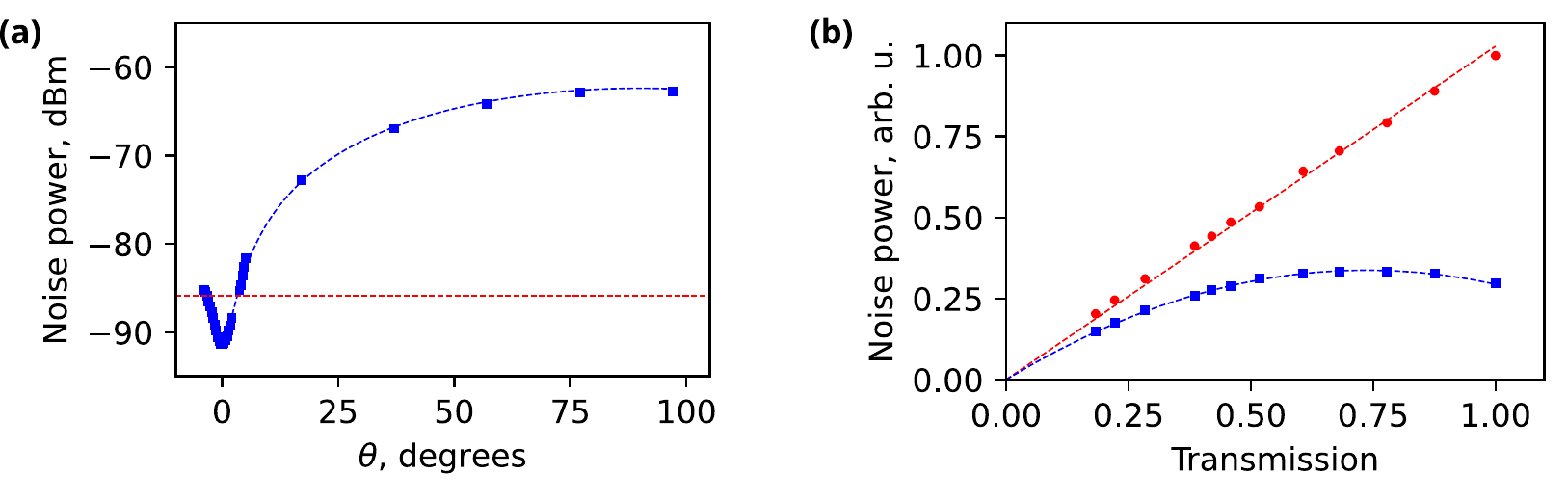}
  \caption{(a) Measured noise power depending on the alignment angle (circles), shot noise level (red dashed line), a fit of formula \eqref{eq:noise_theta} on experimental data (blue dashed line). Pulse duration (FWHM) $200$ fs, pulse energy $160$ pJ. (b) Measured noise power depending on attenuation: shot noise (red squares, dashed is linear fit), squeezing (blue circles, dashed is parabolic fit \eqref{eq:noise_attenuation}). Pulse parameters same as in (a), alignment angle $\theta = 0$.}
  \label{fig:angle}
  \label{fig:attenuation}
\end{figure*}

The shot noise level was measured using the $\hat{S}_0$ variance measurement, it was also verified independently with a setup without the fibre. When the variance of $\hat{S}_1$ was measured, the result depended on the angle of the half-wave plate that changes the angle of the ellipse $\theta$ on the Poincar\'e sphere (Fig. \ref{fig:sphere}). The lowest noise corresponded to the measurement of the squeezed axis of the fibre and defines the squeezing value, while the highest noise corresponded to antisqueezing. The dependence of the noise power on the alignment of the ellipse is shown in Fig. \ref{fig:angle}a, perfectly matching the expected behavior. The maximum observed squeezing was $>5$ dB.

When the beam is attenuated in front of the Wollaston prism, the expected dependence of the noise power on the transmission is given by
\begin{equation}
    \Delta^2_{\mathrm{det}} = T \left[T \Delta^2 + (1 - T) \Delta_{\mathrm{sn}}^2\right],
    \label{eq:noise_attenuation}
\end{equation}
where $\Delta^2_{\mathrm{det}}$ is the detected noise, $T$ is transmission, $\Delta^2$ is the noise of a non-attenuated beam, and $\Delta_{\mathrm{sn}}^2$ is the shot noise at the same power. For a coherent beam, the dependence is linear, because $\Delta^2 = \Delta_{\mathrm{sn}}^2$, and for a squeezed beam it is a parabola pointing downwards. The noise levels of the squeezed beam measured in the experiment demonstrated parabolic dependence, confirming that genuine squeezing was seen (Fig. \ref{fig:attenuation}b).

\begin{figure}[t]
  \centering
  \includegraphics[width=0.9\columnwidth]{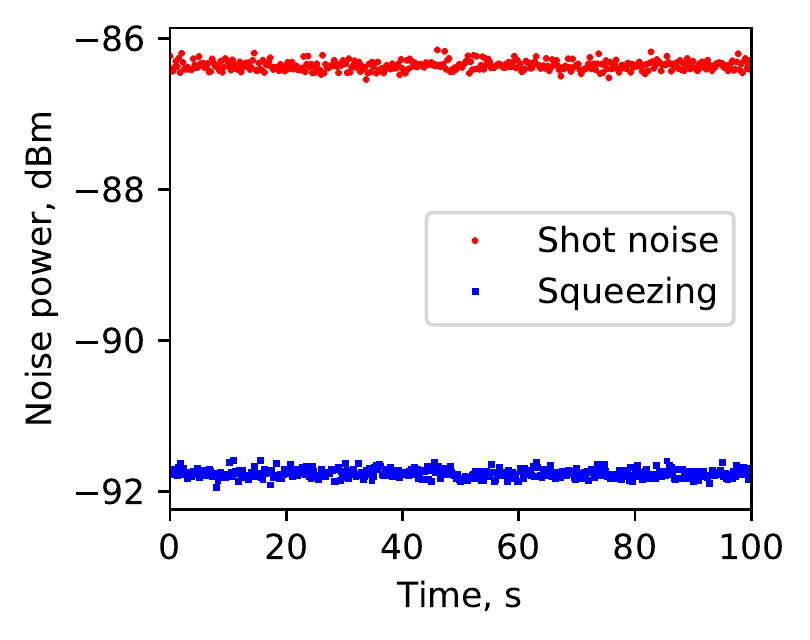}
  \caption{Measured shot noise power (red squares) and squeezing of $>5$ dB (blue circles) over a period of 100 s. Pulse duration (FWHM) $200$ fs, pulse energy $160$ pJ, alignment angle $\theta = 0$.}
  \label{fig:longtime}
\end{figure}

The squeezing generation was very stable without any adjustments, it deteriorated on a time scale of days due to environmental changes. The shot noise and squeezing noise for a period of $100$ s are shown in Fig. \ref{fig:longtime}. 

The losses of the detection setup were estimated to be $12\%$ ($4\%$ from the fibre end, $6\%$ from the optical elements, $2\%$ from the quantum efficiency of the detectors). Additionally, in-fibre losses were estimated to be $8\%$ ($4\%$  from the splice loss and $4\%$ from the non-ideal interference of the two polarization modes). Thus the inferred squeezing value assuming ideal detection setup is $-6.5$ dB, and assuming no losses in the fibre is $-8.4$ dB. We assume that the amount of squeezing in our setup was limited by the length of the fibre, and higher values can be achieved in future work using a longer piece of fibre.

\begin{figure}[t]
  \centering
  \includegraphics[width=0.9\columnwidth]{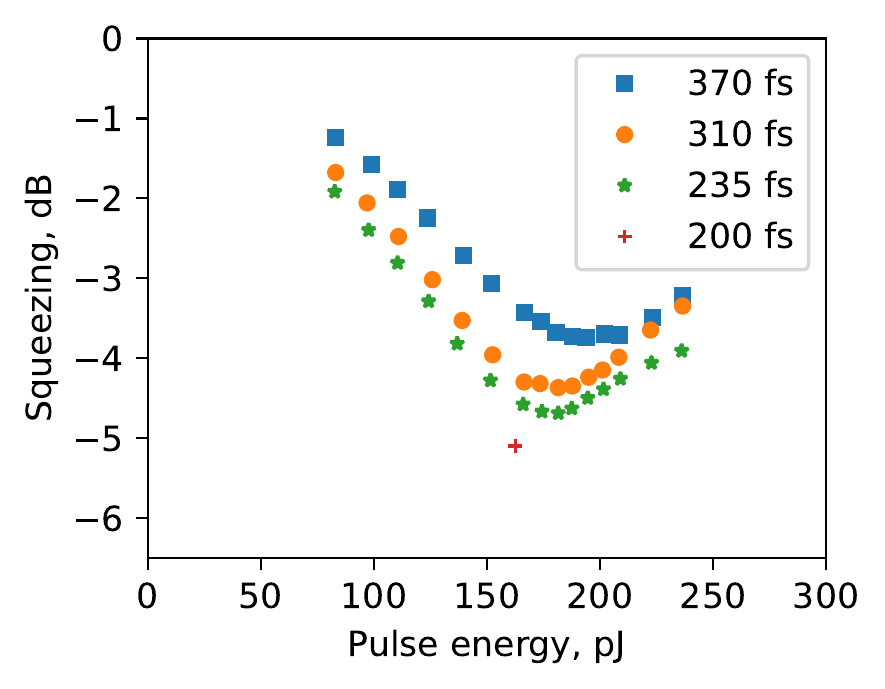}
  \caption{Measured squeezing for different pulse energies for the pulse duration (FWHM) of 200 fs (red plus), 235 fs (green stars), 310 fs (orange circles), 370 fs (blue squares).}
  \label{fig:power_dur}
\end{figure}

The amount of observed squeezing demonstrated strong dependence on the pulse energy (Fig. \ref{fig:power_dur}), similar to what was observed in \cite{corney_2008}. However, in our work, the optimal pulse energy was higher than two soliton energies (one soliton energy in each polarization mode, $105$ pJ in total for $235$ fs FWHM pulse duration). We attribute this behavior to the shorter fibre length so that in our experiment the pulses of similar to soliton powers do not propagate long enough in the fibre to significantly change their shapes. Indeed, the fibre length was only between 1 and 3 soliton periods for the pulse duration used in the experiment. On the other hand, the peak power of higher-energy pulses is higher, which increases the amount of squeezing. This explanation is confirmed by the fact that when the pulse duration is increased, the optimal power is also increased to maintain similar peak power, as opposed to be decreased to maintain soliton shape (Fig. \ref{fig:power_dur}).

\section{Numerical modeling}

To support our experimental results, we performed a series of numerical simulations. In order to calculate squeezing, we followed a phase-space method using the Wigner representation described in detail in \cite{Drummond:01}. The important effects this model includes are the second and third order dispersion, the instantaneous and the delayed (Raman) parts of the nonlinear $\chi^{(3)}$ response. For our fibre we measured and used in the modeling the following relevant parameters: second-order dispersion $\beta_2 = -10.5$ \si{fs^2/mm}, third order dispersion $\beta_3 = 155$ \si{fs^3/mm}, effective nonlinearity $\gamma = 3.0 \cdot 10^{-3}$ \si{1/(W\cdot m)}. We also included linear losses in the fibre and in the detection scheme, which give $20\%$ losses in total.

The previously developed model describes the quantum evolution of pulses in a single spatial mode well.  To model our experiment, it is vital to include the effects arising from pulses propagating in two polarization modes simultaneously. Due to different group velocities in the two modes, the pulses only significantly overlap in time in the first $\sim 40$ cm of the fibre, and the last $\sim 40$ cm of the fibre. However, the interaction between the pulses via the cross phase modulation (XPM) effect cannot be neglected, especially at high pulse energy. The most pronounced effect of the XPM in the initial part of the fibre is that the central frequencies in the spectra of the two polarization modes undergo a transient change \cite{Koenig2002,Mollenauer1991}, resulting in a slight time delay difference at the end of the fibre. This effect can be partially compensated by adjusting the energies of the two pulses, or by adjusting the lengths of the two halves of the fibre. In fact, in the experiment, the lengths of the fibre halves were slightly different to compensate the XPM effect at the optimum pulse energy. Our modeling includes both the XPM effect and the difference in lengths of $3$ cm between the two halves.

\begin{figure}[t]
  \centering
  \includegraphics[width=0.9\columnwidth]{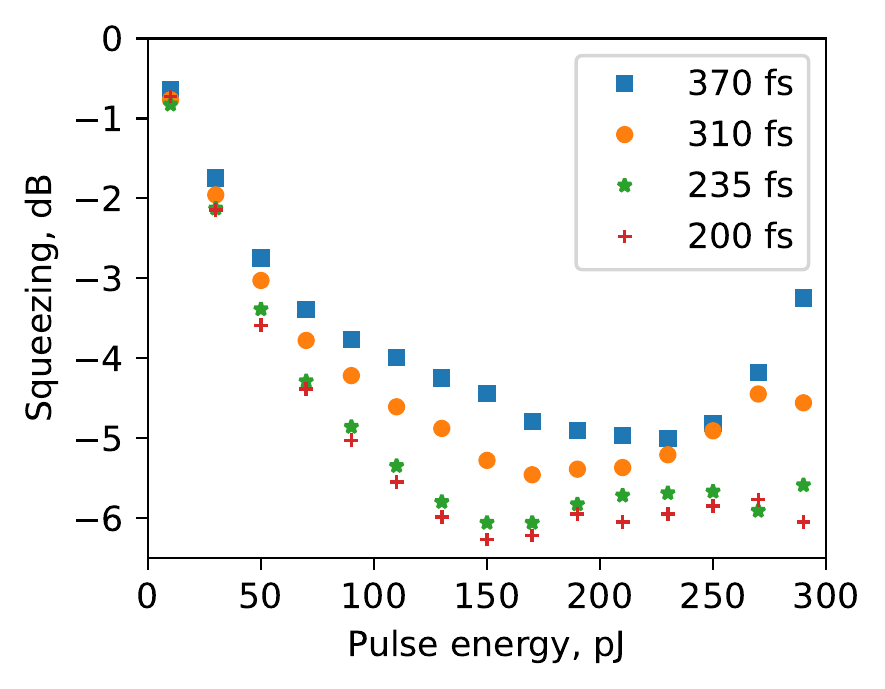}
  \caption{Calculated squeezing for different pulse energies for the pulse duration (FWHM) of 200 fs (red crosses), 235 fs (green stars), 310 fs (orange circles), 370 fs (blue squares).}
  \label{fig:simulation}
\end{figure}

For each data point, we modeled $3000$ Wigner trajectories in phase space using the stochastic nonlinear Schr{\"o}dinger equation, this number of trajectories is enough for $0.1$ dB precision. The results of the modeling are shown in Fig. \ref{fig:simulation}. The results are in good qualitative agreement with the experimental results, demonstrating better squeezing for shorter pulse duration, while the optimum power is larger for the longer pulses. In order to have a better quantitative agreement between the simulation and the experiment, one has to additionally include several effects which are difficult to estimate in the experiment from first principles. One of them is the GAWBS effect, most of which is cancelled due to simultaneous co-propagation of the pulses. However, some residual effect is still present and can decrease squeezing, especially at low pulse energies. Additionally, not included in the modeling is the effect  that at the splice point, each of the two polarization modes of the first half of the fibre may not couple only into one but a little bit and unintended also into the other orthogonal fibre mode of the second half of the fibre, which creates two low energy pulses before and after the main pulse and increases the noise power. In an attempt to explain this residual discrepancy, further simulations are underway.

\section{Conclusion and outlook}

The novel scheme presented here led to a robust and conceptually simple generation of sizeable squeezing 
by using strictly co-propagating beams
The next steps will be systematically exploring the experimental parameters further trying to find the optimum conditions for producing the best squeezing possible with this scheme. This includes variations of parameters --- not only pulse energy and fibre length but also material parameters --- modifying the type of nonlinear material used to fabricate the wave guide \cite{Anashkina2020,ANASHKINA2021a,Anashkina2021b}. 
One type of promising new material is silicon nitride forming wave guides integrated on a chip. The experimental exploration of this material platform in various geometries has already started  \cite{Dutt2015,Dutt:2016,Lipson2020,Zhang2021,Vuckovic2022}. 
For ultrashort pulsed pump light, simulations had found that the eventual reduction in the amount of squeezing for higher pulse energies is caused by Raman noise \cite{corney_2008}. This is confirmed by the simulations reported here. However, recently it was found, that in special fibre geometries the Raman noise may be significantly suppressed \cite{Liu2022}. It will be studied, whether this suppression can also be achieved under the conditions required for squeezing the quantum noise and whether the amount of squeezing can be further improved.   

\section*{Acknowledgments}
This work was supported by the Mega-grant of the Ministry of Science and Higher Education of the Russian Federation, contract No. 075-15-2021-633 and by Russian Foundation for Basic Research, Grant No. 19-29-11032. A.A.S. acknowledges support from the Foundation for the Advancement of Theoretical Physics and Mathematics ``BASIS''.

\bibliography{refs.bib}

\end{document}